\documentclass{article}
\usepackage{spconf,amsmath,graphicx}
\usepackage{booktabs}
\usepackage{tikz}
\usetikzlibrary{calc}
\usepackage{amssymb}
\usepackage{graphicx}
\usepackage{amsmath}

\usepackage{siunitx}
\usepackage{balance}
\usepackage{subcaption}
\usepackage{xcolor}
\usepackage[hang,flushmargin]{footmisc}
\usepackage{verbatim}

\usepackage{algorithm}
\usepackage{algorithmic}

\usepackage{romannum}
\newcommand{\RomanNumeralCaps}[1]{\MakeUppercase{\romannumeral #1}}
\include{pythonlisting}

\title{ACCELERATED INTRAVASCULAR ULTRASOUND IMAGING USING DEEP REINFORCEMENT LEARNING}
\name{
    Tristan S.W. Stevens$^{\star}$, \quad
    Nishith Chennakeshava$^{\star}$,
    }
\secondlinename{
    Frederik J. de Bruijn$^{\dagger}$, \quad
    Martin Pekař\,$^{\dagger}$, \quad
    Ruud J.G. van Sloun$^{\star\dagger}$
    }

\address{$^{\star}$Department of Electrical Engineering, Eindhoven University of Technology, $^{\dagger}$Philips Research}

\begin{document}
%
\maketitle
\begin{abstract}
Intravascular ultrasound (IVUS) offers a unique perspective in the treatment of vascular diseases by creating a sequence of ultrasound-slices acquired from within the vessel. However, unlike conventional hand-held ultrasound, the thin catheter only provides room for a small number of physical channels for signal transfer from a transducer-array at the tip. For continued improvement of image quality and frame rate, we present the use of deep reinforcement learning to deal with the current physical information bottleneck. Valuable inspiration has come from the field of magnetic resonance imaging (MRI), where learned acquisition schemes have brought significant acceleration in image acquisition at competing image quality. To efficiently accelerate IVUS imaging, we propose a framework that utilizes deep reinforcement learning for an optimal adaptive acquisition policy on a per-frame basis enabled by actor-critic methods and Gumbel top-$K$ sampling.
\end{abstract}
\vspace{-2mm}
\section{Introduction}
\vspace{-2mm}
\label{sec:intro}
Minimally invasive vascular interventions are increasingly guided by phased array intravascular imaging. Intravascular ultrasound (IVUS) is a catheter-based imaging modality that enables detailed assessment of the lumen, endothelium and surrounding tissue of blood vessels. Most importantly, it enables the verification of the treatment, generally the placement of one or more stents.

Ultrasound (US) is an affordable, safe and overall convenient medical imaging tool. However, in general, ultrasound images are notoriously difficult to interpret, due to the characteristics of current methods of ultrasound image formation. For IVUS images, this means that the artifacts such as clutter and speckle strongly affect the appearance of small details and of subtle variations in texture. A good image quality facilitates the important distinction between tissues but also e.g. between different types of arterial plaque. It also makes it easier to verify the proper placement of a stent. 

Image quality in phased array IVUS is fundamentally bounded by the information extracted by the adopted pulse-echo imaging sequence. Subsampling of this acquisition sequence to boost frame rates while maintaining image quality is an active area of research in ultrasound in general \cite{wagner2011xampling, huijben2020learning}, but its benefits are especially promising for IVUS as its form-factor highly limits the bandwidth for the raw sensor data. Careful selection of which measurements to sample will not only reduce the amount of data, but subsequently decrease acquisition time and potentially simplify the hardware. Moreover, it allows extra time for dedicated imaging modes, such as spectral Doppler imaging and high-resolution elastography, potentially providing additional clinical information. Furthermore, the reconstruction of the blood vessel is captured in a sequence of frames by the gradual withdrawal of the transducer-bearing catheter. These \emph{pullbacks} are often accompanied by motion artifacts \cite{torbati2019image}. High-frame rate imaging is therefore desirable as it alleviates these adverse motion effects and allows for faster pullbacks thereby reducing the overall procedure time. In a clinical setting, this directly translates to quicker interventions. 

\vspace{-0.5mm}
Model-based approaches towards subsampling of ultrasound data focus on either sub-aperture imaging \cite{karaman1998subaperture} or sparse array designs such as Cantor arrays \cite{liu2017maximally}. These sequences are hand-engineered, static, and not jointly optimized with downstream image reconstruction methods. 
In the past decades, digitization of raw radio-frequency (RF) channel data and the subsequent beamforming required custom signal processing hardware, to achieve real-time performance. Today, the availability of affordable, fast general-purpose high-performance processors opens the way to implement increasingly sophisticated algorithms for real-time ultrasound image formation. This revolution in high-performance computing has spurred the adoption of deep learning (DL) in ultrasound imaging \cite{van2019deep, luijten2020adaptive, hase2020ultrasound}. Naturally, this trend has also led to the use of compressed sensing (CS) in ultrasound imaging \cite{wagner2011xampling, huijben2020learning}.

\vspace{-0.5mm}
Therefore, we propose learning optimal IVUS-specific acquisition-sequencing using state-of-the-art DL methods in a framework we dub as \emph{AiVUS}. The framework learns to optimally extract information from the acoustic scene by adaptive sampling of transducer element pairs. AiVUS poses imaging as an information-gain problem, where the intelligent system measures precisely what is needed to provide at any time the optimal trade-off between imaging properties, disturbing artifacts, and constraints on data rates. 

To that end, AiVUS relies on deep reinforcement learning (DRL). Researchers have employed DRL with great success in a variety of applications, from magnetic resonance imaging (MRI) \cite{bakker2020experimental} and computed tomography (CT) \cite{ghesu2017multi} to radar \cite{stevens2021automated}. Adaptive sampling methods for medical imaging that use DRL have predominantly been developed for MRI \cite{bakker2020experimental, jin2019self}. These frameworks seek to iteratively optimize the measurements taken for a single reconstruction. This means that the RL loop is executed for every measurement necessary to construct the frame. However, we argue this approach does not extrapolate well to IVUS, as the scan time in MRI is orders of magnitude longer compared to ultrasound acquisition  \cite{zbontar2018fastmri}. In contrast to these methods, AiVUS selects all measurements needed for a single IVUS reconstruction concurrently, such that the agent is employed on a per-frame basis. This means that the sampling strategy of the current frame is conditioned on the reconstruction of the previous frame, which benefits the faster IVUS acquisition process. Granted, sufficient temporal correlation between successive frames is assumed. In other words, the current imaging scene is informative enough for the policy to provide valid sampling for the next frame.

Other deep learning subsample strategies (not using DRL), such as Deep Probabilistic Subsampling (DPS) \cite{huijben2020learning} and Active-DPS (A-DPS) \cite{gorp2021active} leverage the so-called `Gumbel-max trick' \cite{maddison2014sampling}, a reparameterization 'trick' to jointly optimize a sampling- and reconstruction-network. DPS learns a fixed sample scheme using a generative sampling model, while A-DPS extends this to an active acquisition framework that conditions the sampling procedure on already acquired data. Both DPS and A-DPS require fully differentiable signal processing pipelines for joint optimization with the sample network. This inherently limits the scope of applications.

\vspace{-2mm}
\section{Methods}
\vspace{-2mm}
\label{sec:methods}
\subsection{IVUS Signal Acquisition}
\label{sec:ivus}
\vspace{-2mm}
Circular-array intravascular ultrasound imaging utilizes uniformly spaced transducers to generate a tomographic reconstruction of the blood vessel wall. Multiple elements eliminate the need for rotating mechanical systems by virtue of an array processing technique known as beamforming, which aims to maximize spatial resolution. During beamforming, focused virtual beams are constructed from unfocused ultrasound pulse-echo signals. The fundamental principle of delay-and-sum beamforming is to arrange the raw data such that the coherent signals are enhanced while the incoherent signals are suppressed through temporal adjustment (time-of-flight correction) and spatial adjustment (apodization). Due to the limited space in the operational domain, severe hardware constraints are imposed on the catheter-based IVUS probe, leaving room for only small circuitry and for only a small number of signal wires. Therefore, each acquisition, corresponding to a transmitting-receiving transducer-element pair, is time-multiplexed and sent over the catheter one at a time. This provokes a trade-off between the image resolution and data acquisition time, which scales linearly with the amount of measured data.

Because of the circular nature of the array, it is advantageous to only combine pulse-echo responses of neighboring transducers. Only elements that significantly contribute to the resulting focused beam are used. For reconstruction of a single IVUS image, the number of measurements is equal to $N = E \times A$, where $E$ is the number of ultrasound elements and $A$ is the sub-aperture size. Each of these $N$ measurements is a vector containing the fast-time RF samples.

\vspace{-4mm}
\subsection{Subsampling}
\label{sec:subsampling}
\vspace{-2mm}
Given the RF channel data $\mathbf{x} \in \mathbb{R}^N$, we are looking for an optimal binary subsampling mask $\mathbf{a}\in\left\{0, 1\right\}^N$ that selects $K \leq N$ measurements from the total of $N$ possible measurements (i.e. element pairs), resulting in an $N / K$ acceleration of the acquisition process. The measured data is given by
\vspace{-2mm}
\begin{equation}
    \mathbf{y} = \mathbf{a} \odot \mathbf{x},
\vspace{-2mm}
\end{equation}
where $\odot$ denotes an element-wise multiplication. Given the current partial acquisition $\mathbf{y} \in \mathbb{R}^N$, which is zero-filled at the non-sampled indices, we would like to design an adaptive subsampling mask $\mathbf{a}$ that maximizes image quality of the next frame. The reconstruction operation $g(\cdot)$ encompasses all signal processing operations for B-mode imaging, after which the IVUS image $\mathbf{s} = g(\mathbf{y})$ follows. A fixed apodization scheme was used for the sake of simplicity. This may require further investigation in future work.

\vspace{-4mm}
\subsection{Reinforcement Learning for Adaptive Subsampling}
\begin{figure}
    \centering
    \includegraphics[scale=0.9]{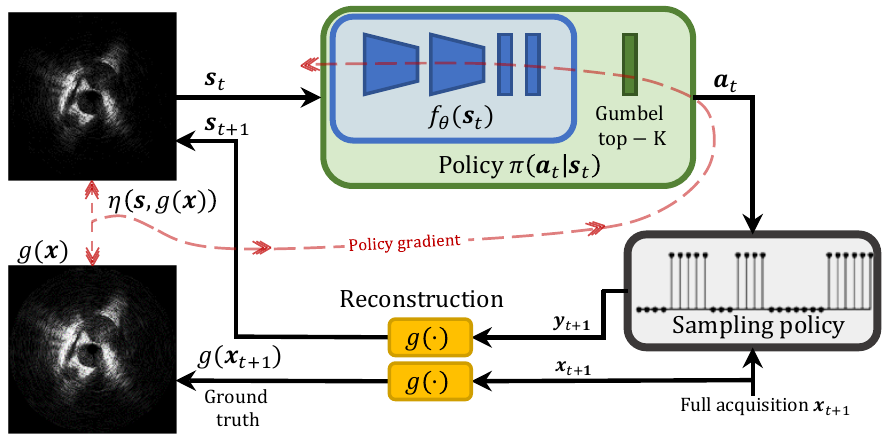}
    \caption{Block diagram of the perception-action-learning loop of the RL agent in an IVUS imaging setting.} 
    \label{fig:rl_loop}
    \vspace{-2mm}
\end{figure}

\vspace{-2mm}
\label{sec:AiVUS}
The model in AiVUS that performs the adaptive subsampling is trained using reinforcement learning (RL). An RL agent takes actions upon observation of its interaction with the environment and can learn to update its behavior accordingly. Transitions in this perception-action-learning loop are modeled as a Partially Observable Markov Decision Process (POMDP). Each step $t$ in the POMDP corresponds to the acquisition and reconstruction of a single IVUS image $\mathbf{s}$. An action $\mathbf{a}_t$ represents a partial acquisition of IVUS data, according to some transmit scheme. The agent produces a policy $\pi(\mathbf{a}_t|\mathbf{s}_t)$ over actions given the current reconstruction $\mathbf{s}_t$ and is incentivised by a reward signal $r_t(\mathbf{s}_t, \mathbf{a}_t)$ that captures the reconstruction quality of $\mathbf{s}$ with respect to the ground truth image $g(\mathbf{x})$. The goal of the agent is to maximize the expected discounted cumulative reward $J_\pi = \sum_{t=0}^{T-1} \mathbb{E}_{(\mathbf{s}_t, \mathbf{a}_t)\sim \pi}\left[\gamma^t r(\mathbf{s}_t, \mathbf{a}_t)\right]$, where the episodes of length $T$ are indexed by $t\in\left[0, T-1\right]$. The discount factor $\gamma$ weighs the reward scores such that more emphasis is placed on immediate rewards. In order to meet the high dimensional state and action spaces, an actor-critic framework is leveraged for training the RL policy \cite{lillicrap2015continuous, haarnoja2018soft}. Actor-critic methods extend the use of an action-value function $Q$, modeled by the critic network, with an explicit definition of the policy, i.e. the actor network $\pi$ parameterized by $\theta$. The critic network $Q$ parameterized by $\phi$ can be iteratively updated through stochastic gradient descent using the Bellman error \cite{lillicrap2015continuous, haarnoja2018soft}:
\vspace{-1mm}
\begin{equation}
    J_Q(\phi) = \mathbb{E}_{(\mathbf{s}_t, \mathbf{a}_t, \mathbf{s}_{t+1}) \sim \mathcal{D}}
    \left[ 
        \big( Q_\phi(\mathbf{s}_t,\mathbf{a}_t) - \hat{Q}(\mathbf{s}_t, \mathbf{a}_t) \big)^2
    \right],
    \label{eq:MSBE}
\vspace{-1mm}
\end{equation}
where $\mathcal{D}$ is the Replay Memory, which stores state transitions sampled from the environment. The target value $\hat{Q}$ given by
\vspace{-1mm}
\begin{equation}
    \hat{Q}(\mathbf{s}_t, \mathbf{a}_t) = r(\mathbf{s}_t, \mathbf{a}_t) + \gamma Q_{\phi'} (\mathbf{s}_{t+1}, \pi_{\theta'}(\mathbf{s}_{t+1})),
    \label{eq:bellman_eq}
\vspace{-1mm}
\end{equation}
defines the recursive relation of the action-value function through temporal difference (TD) learning. The target is computed with separate target networks parameterized by $\phi', \theta'$ to improve stability of learning \cite{lillicrap2015continuous}. The actor network is updated through gradient ascent using the policy gradient
\vspace{-1mm}
\begin{equation}
    \nabla_\theta J_\pi(\theta) \simeq \mathbb{E}_{\mathbf{s}_t \sim \mathcal{D}, \mathbf{e}_t \sim \text{Gumbel}} 
    \left[ 
        \nabla_\theta Q_\phi (\mathbf{s}_t, \pi_\theta(\mathbf{s}_t)) 
    \right],
\vspace{-1mm}
\end{equation}
such that the expected return, modeled by $Q_\phi(\cdot)$, is maximized. Current actor-critic methods such as Deep Deterministic Policy Gradient (DDPG) \cite{lillicrap2015continuous} and Soft Actor-Critic (SAC) \cite{haarnoja2018soft} do not directly support multiple action selection. In the IVUS setting, we cannot afford running the network iteratively to generate multiple actions for a single reconstruction, due to the high pulse-repetition interval. Multiple actions can be modeled by a larger underlying unstructured distribution such that the policy only has to sample a single action. However, because the number of parameters of a distribution over an $N$-choose-$K$ problem scales factorially, modeling the underlying categorical distribution using the policy is intractable. Therefore, AiVUS leverages Gumbel top-$K$ sampling, which allows exact sampling of $K$ actions simultaneously by a single pass of the policy network \cite{kool2019stochastic}. Gumbel top-$K$ sampling is equivalent to sampling $K$ times without replacement from the same categorical distribution. In this way, the logits of the policy network represent only a single categorical distribution of size $N$ from which we take $K$ actions. Gumbel Top-$K$ sampling is a generalization of the Gumbel-max trick \cite{maddison2014sampling} which perturbs the unnormalized logits with i.i.d. Gumbel noise samples
\vspace{-1mm}
\begin{equation}
    \mathbf{e}_t = -\log \left(-\log(U)\right) * \sigma \sim \text{Gumbel}(0, \sigma),
\label{eq:gumbel_noise}
\vspace{-1mm}
\end{equation}
where $U\in[0, 1]$ is sampled from a continuous uniform distribution. The Gumbel samples are sampled with scaling term $\sigma$. The sampling procedure is computed as follows:
\vspace{-1mm}
\begin{equation}
    \mathbf{a}_t = \mathrm{topK}(f_\theta(\mathbf{s}_{t}) + \mathbf{e}_t),
\vspace{-1mm}
\end{equation}
where $f_\theta$ is a deterministic function of the state whose logits represent the unnormalized log-probabilities over actions, from which $\pi_\theta$ follows after top-$K$ sampling. The non-differentiable $\text{argmax}(\cdot)$ operation (used to compute top-$K$) can be relaxed with reparameterizable subset sampling using $\text{softmax}(\cdot)$ \cite{xie2019reparameterizable}, to permit error backpropagation through $f_\theta(\cdot)$, similar as done in \cite{huijben2020learning}. Intuitively, we can scale the Gumbel noise with $\sigma$ to sample more uniform actions, which in return ensures exploration. During training we use an annealing scheme for $\sigma$, starting with a relatively high value ($\sigma=2$), and reducing it until reaching a set minimum value ($\sigma=0.2$). In practice, this annealing approach led to more stable training compared to a fixed setting of $\sigma=1$. During evaluation, we disable the Gumbel noise and take the top-$K$ largest values of the policy distribution. A schematic overview of the full signal path is shown in Fig.~\ref{fig:rl_loop}.

\vspace{-2mm}
\section{Experiments}
\vspace{-2mm}
\label{sec:experiments}
In this work, a prototype capacitive micromachined ultrasonic transducer (CMUT) IVUS sensor with a 114 element circular phased-array (Philips Research \cite{zangabad2021real}) is used to provide the raw RF data for training and verification of the models. We demonstrate our framework on three different datasets, increasing in complexity. Namely, simulated wire targets, wire phantoms and \emph{in-vivo} data from a porcine model. The data is recorded in raw channel RF format and after beamforming each frame contains 456 scanlines with 520 samples along the penetration depth of \SI{8}{mm}. We consider episodic environments with fixed length $T=10$.
\begin{figure*}
\centering
\begin{minipage}[b]{.6\textwidth}
    \includegraphics[trim=50 14 50 15, clip, scale=0.75]{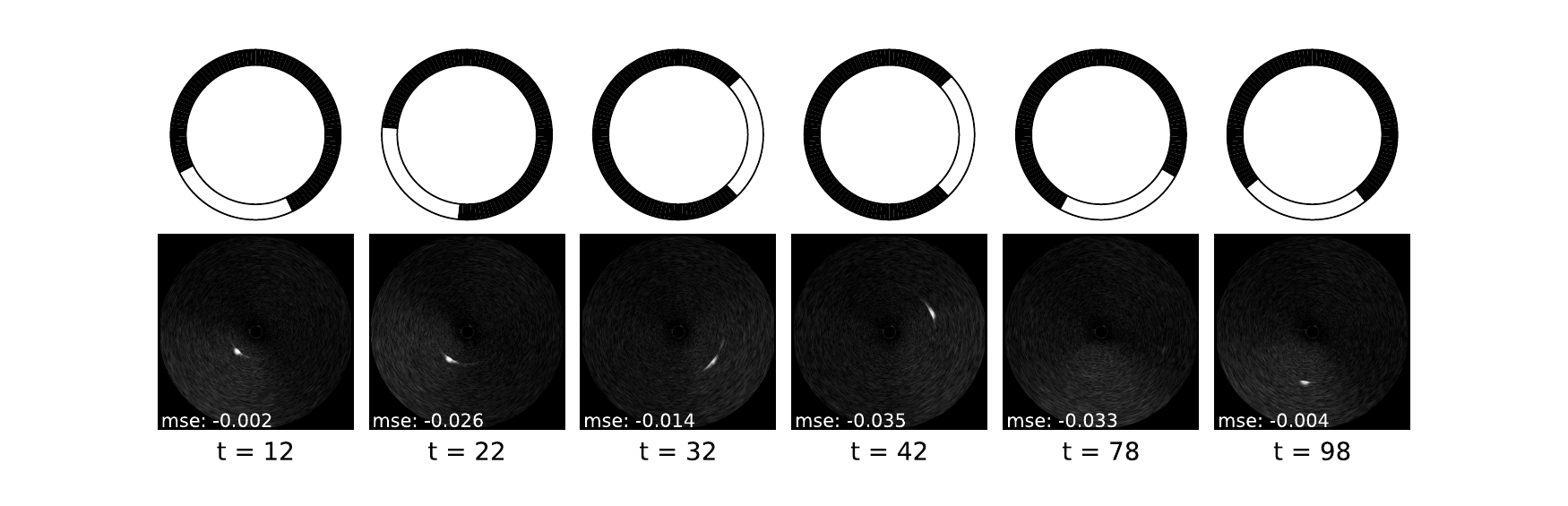}
    \caption{Six successive wire phantom frames constructed using AiVUS. The agent's action is displayed in the top row, where the circle represents the elements in the transducer array. Actions are displayed such that black represents 1 receiving element and white $A$ receiving elements, for each transmitting element $\in \left\{1, \ldots, E\right\}$.}
    \label{fig:snapshots}
\end{minipage}\quad
\begin{minipage}[b]{.38\textwidth}
    \includegraphics[trim=10 13 0 10, clip, scale=0.88]{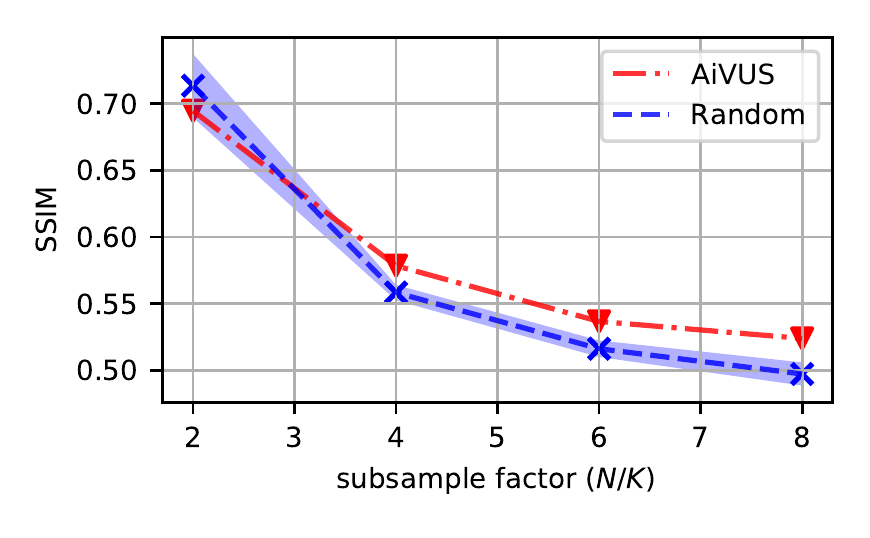}
    \caption{SSIM performance on the \emph{in-vivo} test data for both learned (AiVUS) and random sampling strategies. Average and standard deviations are computed over five seeds.}
    \label{fig:pareto_curve}
\end{minipage}
\end{figure*}
\vspace{-2mm}
\subsection{Simulated Wire Target}
\vspace{-2mm}
\label{sec:sim_wire_target}
The first dataset is created by simulating moving wire targets that revolve around the center at a random distance. We test if AiVUS can learn an adaptive policy and track the targets. In this specific case, it is very clear what kind of policy is expected, i.e. sample the channels that originate from transducers that are in close vicinity to the wire targets. The policy network is trained on 200 frames and evaluated on 100 test examples. A mean squared error (MSE) is used as reward:
\begin{equation}
\vspace{-2mm}
    r_t = \eta_{MSE} = -||\mathbf{s}_{t} - g(\mathbf{x}_t))||_2^2.
\label{eq:rt_sim_target}
\vspace{-2mm}
\end{equation}
\vspace{-6mm}
\subsection{Wire Phantom}
\label{sec:wire_phantom}
\vspace{-1mm}
To bridge the gap from simulated data to more realistic US data, we subject the DRL agent to images of wire phantoms. It allows to assess the performance of the agent in a more realistic environment which encompasses interference and noise as well as more subtle US features such as reverberation and ringdown. In total 15 sequences with 8767 frames are recorded, of which 20\% are randomly assigned to the test set. For the reward calculation, \eqref{eq:rt_sim_target} is used, however, the images are filtered using straightforward thresholding to reduce background noise and interference. We found this step was useful for constructing a consistent reward signal.
\vspace{-1mm}
\subsection{In-vivo data}
\vspace{-1mm}
\label{sec:in_vivo}
Although the wire target experiment shows proof of concept of AiVUS, experiments on \emph{in-vivo} data are necessary to show the framework has potential for clinical relevance. The prior acquired dataset consists of eight different pullbacks from a porcine model with 8679 frames in total of which 20\% is randomly assigned as test set. Each pullback is done at approximately \SI{1}{mm/s} for a length of \SI{30}{mm}. The structural similarity index measure (SSIM) is used as opposed to the MSE for training the policy network. Furthermore, we observe that the speckle, which is the granular appearance inherent to the ultrasound imaging, has a significant effect on the image quality metric. Different sample schemes lead to small shifts in the speckle texture which translate to the reward score. Slight changes in the speckle texture do not positively contribute to the visual appearance as the speckle pattern appears random. To promote sharp reconstruction of local features, we add an adversarial loss by adopting a discriminator network $D_\psi$. The resulting reward score is a combination of the SSIM score and discriminator loss $L_{D_\psi}$ :
\vspace{-2mm}
\begin{equation}
    r_t = \eta_{SSIM} - \lambda_D \mathcal{L}_{D_\psi},
\label{eq:rt_invivo}
\vspace{-2mm}
\end{equation}
where $\lambda_D$ weights the discriminator loss to the image quality metric and was empirically set to \num{1e-5}. To further reduce the effect of the speckle, an anisotropic diffusion filter is applied to both state and ground truth images before the reward calculation.

\begin{table}[t]
    \centering
    \caption{Quantitative results on the test sets of all three experiments with subsampling factor $N/K=4$, comparing a random agent (\RomanNumeralCaps{1}) with a trained agent (\RomanNumeralCaps{2}) (AiVUS).}
    \hspace{-0.3cm}  
    \begin{tabular}{l l l l l l l l l}
    \toprule
    \textbf{Dataset}    & \multicolumn{2}{c}{Sim.} & \multicolumn{2}{c}{Phantom} & \multicolumn{2}{c}{In-vivo}\\
    \cmidrule(rl){2-3} \cmidrule(rl){4-5} \cmidrule(rl){6-7}
    \textbf{Agent}      & \RomanNumeralCaps{1} &\RomanNumeralCaps{2} & \RomanNumeralCaps{1} &\RomanNumeralCaps{2} & \RomanNumeralCaps{1} &\RomanNumeralCaps{2}  \\[0.1cm]
    \midrule
    MSE  $\downarrow$   & 3.25      & 1.42        & 0.067       & 0.047   & 0.078       & 0.070      \\
    MAE  $\downarrow$   & 0.069     & 0.034       & 0.169       & 0.133   & 0.211       & 0.200      \\
    PSNR $\uparrow$     & 44.33     & 49.95       & 61.59       & 64.48   & 59.43       & 59.90     \\
    SSIM $\uparrow$     & 0.996     & 0.998       & 0.308       & 0.447   & 0.552       & 0.578     \\
    \bottomrule
    \end{tabular}
    \label{tab:results}
    \vspace{-2mm}
\end{table}
\vspace{-3mm}
\section{Results}
\vspace{-1mm}
In Table~\ref{tab:results} we report test scores of the final reconstructions using four common image quality metrics. Namely, the mean-squared error (MSE), mean-absolute error (MAE), peak signal-to-noise ratio (PSNR) and the structural similarity index measure (SSIM). The trained DRL agent is compared to an agent which randomly samples from the transducer element pairs with the same subsampling factor. AiVUS outperforms a random agent on all four metrics. Fig.~\ref{fig:snapshots} shows a sequence of successive frames from the wire phantom dataset alongside the sampling actions that were used to facilitate the reconstruction. AiVUS is able to show adaptivity by increasing the aperture size of specific transducers that are located near the target and tracking the targets from frame to frame. The agent demonstrates it can exploit knowledge of the current frame to provide an optimal acquisition sequence for the next reconstruction. For the \emph{in-vivo} data, a comparison of the SSIM scores for different subsampling factors, both for random and learned sampling by AiVUS, is visualized in Fig.~\ref{fig:pareto_curve}. For higher subsampling factors, the learned strategy is preferred, while for a subsampling factor of two there is almost no performance difference.

\section{Conclusion}
\label{sec:conclusion}
In this paper, we have proposed a deep reinforcement learning framework that learns subsampling policies for IVUS on a per-frame-basis using actor critic methods. The framework allows for any non-differentiable reconstruction method and quality metric. We have demonstrated that AiVUS is able to navigate in controlled IVUS environments with high dimensional state and action spaces. AiVUS outperforms a random agent using a learned acquisition strategy. In future work, we foresee this framework can be put to use for control of other ultrasound transmit parameters.
\vfill
\bibliographystyle{IEEEbib}
\bibliography{refs}

\end{document}